\documentclass[a4paper,11pt]{article}
\usepackage{jcappub} 
\usepackage{lineno}

\usepackage{aas_macros}
\usepackage{amsmath}
\newcommand{\mc}[1]{\mathcal{#1}}
\newcommand{\ve}[1]{\boldsymbol{#1}}

\arxivnumber{1234.56789} 
\title{\boldmath Augmented Correlation Functions for Spectroscopic Galaxy Surveys}

\author[1,2]{D. Bianchi}
\affiliation[1]{Dipartimento di Fisica ``Aldo Pontremoli'', Universit\`a degli Studi di Milano,\\ Via Celoria 16, I-20133 Milano, Italy}
\affiliation[2]{INAF-Osservatorio Astronomico di Brera\\ Via Brera 28, I-20122 Milano, Italy}

\emailAdd{davide.bianchi1@unimi.it}

\abstract{Galaxy redshift surveys encode a wealth of information generated by nonlinear gravitational evolution, galaxy bias, and redshift-space distortions, only part of which is accessible through standard two-point statistics. Motivated by the need for flexible and computationally efficient alternatives, we introduce the augmented correlation function, a general framework in which an arbitrary transformation of the galaxy field defines additional ``latent'' dimensions that extend the standard two-point correlation function and isolate clustering properties averaged out in conventional analyses.
As a proof of concept, we study a latent variable constructed from the pairwise gradient of the inverse Laplacian of the galaxy density field, showing that the resulting statistics naturally distinguish clustering regimes associated with infalling and outflowing pairs.
Using Fisher forecasts based on $z=1$ halo catalogues from the Quijote simulations within $\nu\Lambda\mathrm{CDM}$ cosmology, we find that the augmented correlation systematically yields tighter constraints on all cosmological parameters considered. Although these improvements should be regarded as indicative given the exploratory nature of the analysis and the limitations of Fisher forecasts and simulations, our results demonstrate the potential of augmented correlations as a flexible framework for extracting additional information from spectroscopic galaxy surveys.}

\begin{document}
\maketitle
\flushbottom

\section{Introduction}
\label{sec:intro}

Large-scale galaxy redshift surveys have become one of the primary tools for precision cosmology, enabling stringent constraints on the expansion history of the Universe and the growth of cosmic structure.
Most of the cosmological information extracted from these surveys relies on traditional two-point statistics, such as the galaxy power spectrum or correlation function, whose interpretation is well understood within the framework of perturbation theory. 
However, the exceptional statistical precision of modern spectroscopic surveys (e.g. \cite{desi2016a, desi2016b, euclid2025}) opens new avenues for extracting cosmological information beyond what can be probed with traditional two-point statistics alone.   

A natural path forward is to incorporate higher-order correlation functions (e.g. \cite{gil-marin2017, slepian2017, hou2023, ivanov2023, damico2024, novell-masot2025, chudaykin2025}).
In fact, ongoing and forthcoming surveys aim to complement standard two-point analyses with three- and even four-point statistics. Nevertheless, these extensions are challenging on multiple fronts.
From a theoretical perspective, higher-order statistics require increasingly complex perturbative modelling, with a rapidly growing number of contributions and nuisance parameters.
On the observational side, the accurate treatment of survey systematics, selection effects, and redshift-space distortions becomes significantly more demanding as the order of the statistic increases.
Finally, the computational cost associated with estimating these observables, their covariance matrices, and the corresponding likelihoods scales steeply with statistic order, posing practical limitations for their routine inclusion in large-scale survey analyses.

This has motivated growing interest in alternative clustering statistics.
Designed to probe complementary aspects of the density field and its environment-dependent evolution, these methods can complement standard analyses, improve robustness to modelling uncertainties, and potentially unlock additional cosmological information beyond that accessible with conventional two-point statistics alone, thus offering a compelling path forward for maximizing the scientific return of current and next-generation spectroscopic surveys. 
A key ingredient enabling and supporting this new effort is the rapid advancement of computational resources and machine-learning techniques, which make it possible to efficiently build emulators from suites of high-fidelity N-body simulations, thus avoiding the reliance on analytic perturbative modelling of the target statistic (e.g. \cite{nishimichi2019, hahn2023, cuesta-lazaro2024, saez-casares2024}).

Representative examples include the wavelet scattering transform, which encodes multiscale non-Gaussian features in a stable hierarchical representation (e.g. \cite{mallat2012, valogiannis2022, valogiannis2024, regaldo-saint2024}), Minkowski functionals, which characterize the topology and morphology of the density field beyond low-order moments (e.g. \cite{mecke1994, fang2017, liu2022, jiang2024}), k-nearest-neighbour (kNN) statistics, which probe the local density and geometric structure of the galaxy distribution through neighbour distances (e.g. \cite{banerjee2021, yuan2023}), and minimum spanning trees, which trace the connectivity and filamentary structure of the cosmic web (e.g. \cite{barrow1985, alpaslan2014, naidoo2020, naidoo2022}).

Another class of alternative statistics is based on applying two-point statistics to suitably transformed versions of the galaxy distribution.
Among these, void-based observables, such as the void–galaxy cross-correlation function, have emerged as powerful probes of both geometry and gravity because of the relative dynamical simplicity of cosmic voids and their sensitivity to redshift-space distortions (RSD; \cite{kaiser1987}), Alcock–Paczynski (AP; \cite{alcock1979}) effect and modified gravity (e.g. \cite{lavaux2012, hamaus2014, cai2016, nadathur2019, zhao2022}).
Similarly, density-split statistics, which measure galaxy clustering conditioned on the local density environment, have been shown to enhance sensitivity to cosmological parameters and galaxy bias by accessing higher-order information in a controlled manner (e.g. \cite{abbas2006, paillas2021, paillas2023, paillas2024}).
Closely related in spirit are marked correlation functions, in which each galaxy is assigned a weight, or “mark,” based on some local property such as density or luminosity.
By selectively enhancing or suppressing contributions from different environments, marked statistics can amplify specific clustering features, and have therefore been extensively studied as probes of galaxy bias, environmental effects, and modified gravity (e.g. \cite{white2016, armijo2018, hernandez2018, aviles2020, massara2021, karcher2025}).
Another promising technique is density-field reconstruction, originally developed to sharpen the baryon acoustic oscillation (BAO) feature by partially reversing non-linear gravitational evolution \cite{eisenstein2007}. More recent developments extend reconstruction beyond BAO applications, demonstrating that appropriately reconstructed density fields can improve constraints on both late-time cosmology and primordial physics, while also providing a natural framework for combining standard and non-standard summary statistics (e.g. \cite{wang2022, wang2023}).

Since all of these statistics depend, to varying degrees, on transformations of the galaxy field, it is natural to seek a flexible framework in which different transformations can be easily incorporated and explored.
To this end, we introduce a new statistic, which we term the {\it augmented correlation}.
The key idea is to compress any transformation of the field into a new variable defined at any pair of points in space, effectively promoting the standard two-point correlation function to a higher-dimensional statistic.
This latent variable retains information about the clustering that is not completely captured by pairwise separations and would therefore be washed out when averaging to obtain the traditional two-point correlation function.

Broadly speaking, the augmented correlation can be viewed either as a generalization of the marked correlation function where, instead of weighting galaxies, the information encoded in the  field transformation is used to construct one or more latent dimensions, or as an extension of density splits to more general, pair-dependent transformations (see Section~\ref{sec:model} for further details).

Of course, not every transformation carries useful cosmological information; to be informative, it must be connected to a physically relevant process.
The most extensively studied example in the literature is the galaxy density field itself smoothed over a characteristic scale.
Here, as a proof of concept, we focus instead on pairwise infall, estimated from the gradient of the inverse Laplacian of the galaxy distribution.
This choice is motivated by the fact that peculiar velocities generate anisotropies in redshift space through coherent infall and virial motions, while the AP effect introduces a distinct anisotropic signal of purely geometric origin.
Since AP distortions depend only on positions, whereas infall directly traces peculiar velocities, pairwise infall (or a suitable proxy for it) provides a natural candidate to help disentangle the two effects and therefore a compelling test case for our framework.

To assess the information content of the resulting augmented statistic relative to the traditional correlation function, we carry out a standard Fisher analysis based on a large suite of cosmological simulations.

The paper is organised as follows. In Section~\ref{sec:model} we introduce the model, while Section~\ref{sec:fisher} describes the simulations and the procedure used to obtain the Fisher forecasts. The results are presented in Section~\ref{sec:results}, and we conclude in Section~\ref{sec:conclusions}.

\section{The augmented correlation}\label{sec:model}

The traditional two point correlation function is defined as $\xi = \langle \delta(\ve{x}_1) \delta(\ve{x}_2) \rangle$, where the brackets indicate ensemble average.
In practice, we will always assume ergodicity and substitute ensemble with volume average.
The density contrast is defined, as usual, by
\begin{equation}\label{eq:delta}
\delta = \frac{\rho}{\langle \rho \rangle} - 1 \ ,
\end{equation}
where $\rho$ denotes the galaxy density.
We define the augmented 2-point statistics $\xi_A$ as
\begin{equation}\label{eq:def_general}
 \xi_A(\ve{x}_1, \ve{x}_2, \lambda) = \frac{\left\langle \delta(\ve{x}_1)\delta(\ve{x}_2) \delta_D \left\{\lambda - K_p\left[\ve{\psi}(\ve{x}_1), \ve{\psi}(\ve{x}_2)\right] \right\} \right\rangle} {\left\langle \delta_D \left\{\lambda - K_p\left[\ve{\psi}(\ve{x}_1), \ve{\psi}(\ve{x}_2)\right]\right\} \right\rangle} , \
\end{equation}
where the field $\ve{\psi}$ is a function, in general non local, of the density contrast $\delta$ and $K_p : V \times V \rightarrow \mathbb{R}$ is a kernel that maps any pair of vectors into a new variable $\lambda$.
In general, the kernel is also a function of the coordinates $\ve{x}_1$ and $\ve{x}_2$; this dependence has been suppressed above for brevity.
As usual, $\delta_D$ denotes the Dirac delta.

This formulation allows for the choice of any physically motivated (and computationally tractable) auxiliary field $\ve{\psi}$ and ``pooling'' kernel $K_p$, thus providing a useful common ground to explore the many different options.  
More in general, the auxiliary field need not be a vector field; it may instead be a tensor field of arbitrary rank, including the special case of a scalar field. Likewise, one may introduce multiple auxiliary variables, $\lambda_1, \dots, \lambda_n$ with $n>1$, by employing a higher-dimensional pooling kernel $K_p : V \times V \rightarrow \mathbb{R}^n$.

In a nutshell, the idea is to expand the usual correlation function into an arbitrary ``latent'' space, described by the coordinate $\lambda$.
For discrete $\lambda$, the process may be viewed as a sample split.
However, unlike other conventional approaches, such as density-splits, we retain the flexibility to perform the split at the pairwise level (see Section~\ref{sec:results} for further details).

Examining Eq.~\ref{eq:def_general} more closely, we see that $\xi_A$ corresponds to the ratio of the density-contrast–weighted distribution of $\lambda$ at a fixed location in pair space (numerator) to its unweighted counterpart (denominator).
The role of the denominator is to ensure that all possible $\lambda$ values are ``treated equally'', i.e. $\xi_A$ is not automatically suppressed for rare values of $\lambda$.
Clearly, $\xi_A$ is defined only on the support of the denominator.
This introduces neither ambiguities nor information loss, because the construction ensures that the numerator is nonzero only when the denominator is nonzero.

In principle, a more general definition could introduce a hyperparameter $\alpha$ that raises the denominator to a power (allowing $\alpha=0$), thereby reweighting different $\lambda$ values, analogous to the $q$ parameter in the scattering transform (e.g. \cite{valogiannis2022}).
Throughout this work, we consider only $\alpha=1$. For different choices of $\alpha$, some of the properties of $\xi_A$ derived below would either fail to hold or would need to be suitably adjusted.

Alternative clustering statistics are typically designed to access higher-order, non-Gaussian information that cannot be captured by conventional 2-point analyses. The class of statistics introduced here is no exception.
However, it is important to emphasize the often overlooked point that part of the enhanced constraining power may simply reflect a less aggressive compression of genuine 2-point information.
In other words, in the hypothetical case of a noiseless $\xi(s, \mu)$ measured with arbitrarily small bins and no multipole compression, the resulting information gain could be diminished (see Section~\ref{sec:results} for further discussions).

Following standard practice, we construct an estimator for the augmented correlation by supplementing the galaxy catalogue with a set of unclustered tracers spanning the same survey volume, hereafter referred to as ``randoms''.
We further adopt the conventional $(s,\mu)$ parametrization,
where $s$ denotes the magnitude of the separation vector $\ve{s} = \ve{x}_1 - \ve{x}_2$, and $\mu$ is the cosine of the angle between $\ve{s}$ and the line of sight.
Under homogeneity and isotropy broken only by line-of-sight effects, these exhaust the relevant two-point degrees of freedom.
Substituting Eq.~\ref{eq:delta} into Eq.~\ref{eq:def_general} it can be shown  (Appendix \ref{app:estimator}) that an estimator for $\xi_A$ is given by
\begin{equation}\label{eq:estimator}
    \hat{\xi}_A(s, \mu, \lambda) = \frac{DD(s, \mu, \lambda) - 2DR(s, \mu, \lambda) + RR(s, \mu, \lambda)}{RR(s, \mu, \lambda)} \ ,
\end{equation}
where $DD$, $DR$ and $RR$ are the galaxy-galaxy, galaxy-random and random-random pair counts, respectively, in arbitrary
bins of $s$, $\mu$ and $\lambda$, normalised by the corresponding total number of pairs.
This expression is a straightforward extension of the Landy–Szalay estimator~\cite{landy1993} that incorporates the auxiliary variable.

\subsection{Quantile decomposition of the latent variable}\label{sec:quantiles}

It is convenient to introduce the variable $q \in (0,1)$, which replaces $\lambda$ via the following change of variables,
\begin{align}
    s &= s \\
    \mu &= \mu \\
    \lambda &= \mc{C}^{-1}(q | s, \mu) \ .
\end{align}
$\mc{C}$ is a cumulative function defined by $\mc{C}(\lambda|s,\mu) = \int_{-\infty}^{\lambda} d\lambda' \ \mc{P}_V(\lambda'|s,\mu)$, where $\mc{P}_V$ is the volume weighted distribution of $\lambda$, i.e. the quantity at the denominator in Eq. \ref{eq:def_general}. 
The transformation is well defined for any distribution that is nonzero almost everywhere over its domain, a condition that should be satisfied for any reasonable choice of the auxiliary field.
Despite its seemingly elaborate form, this is simply a way to write $\xi_A$ as a continuos function of the quantiles of $\lambda$, which are obtained when $q$ is evaluated in linear bins.

For practical purposes, one simply needs to determine which $\lambda$ bin edges \\ $(\lambda_0, \lambda_1),(\lambda_1, \lambda_2),\dots,(\lambda_{N_q-1},\lambda_{N_q})$ divide the random pair counts $RR(s,\mu,\lambda)$ into $N_q$ equal parts, where $N_q$ is, by construction, the number of quantiles.
If the random counts in Eq.~\ref{eq:estimator} are expressed in terms of these quantiles, they become independent of the latent variable at any separation, that is, $RR(s,\mu,q) = RR(s,\mu)/N_q$. Conversely, the latent variable $\lambda$, or more precisely $q$, remains explicitly present in the data and cross counts, since the bin edges that divide $RR$ into equal parts do not, in general, do the same for $DD$ and $DR$.
   
Switching to quantiles is not strictly necessary. The choice of whether to work with the original latent variable $\lambda$ instead ultimately depends on how $\lambda$ is defined and on the specific goals of the analysis.
For example, quantiles are invariant under a global rescaling of the amplitude of $\lambda$.
If the amplitude is expected to carry relevant information, then quantiles may not be the optimal choice.

On the other hand, quantiles provide a natural way to distribute the noise more evenly. More generally, we argue that they help make the augmented correlation less sensitive to number-density, or shot-noise–like, effects. While, in principle, such effects contain information and could tighten cosmological constraints, in practice they are poorly understood and difficult to model, even in high-fidelity simulations. This issue affects, to varying degrees, all alternative clustering statistics, and having the ability to remove or mitigate these effects is therefore desirable.
At this stage, however, these considerations remain speculative and require further testing.

One useful property of the quantile description is that, since the $RR$ counts are independent of the latent variable, the standard two-point correlation function is simply the algebraic mean of the different $q$-slices of the augmented correlation,
\begin{equation}\label{eq:xi_mean}
\xi(s,\mu) = \frac{1}{N_q}\sum_{i=0}^{N_q-1} \mc{Q}_i(s,\mu) \ ,
\end{equation}
where $\mc{Q}_i(s,\mu)$ is the $i$-th quantile.

The use of quantiles is not new in this type of analysis; for instance, they are commonly employed in density-split studies.
The practical difference in our case is that the quantiles are evaluated at each separation.
This helps regularize the statistics when the distribution of the latent variable varies significantly with separation, as in the case discussed below.

When considering $n>1$ latent variables, the definition of the quantiles is no longer unique, as different choices can be made to partition the $n$-dimensional space into equal-probability regions.
If the $n$ variables are independent, such a partition can be naturally constructed by combining the individual quantiles of each variable. 
Otherwise, the partitioning scheme should be informed by the shape of their joint distribution.

Equation~\ref{eq:xi_mean} provides a straightforward way to show that the augmented correlation contains the standard correlation function $\xi$ and, therefore, can only add information to it (at least in the noiseless limit).
A natural question that follows is whether one can organize the full information content such that $\xi$ explicitly appears as one of the $N_q$ statistics, while the remaining $N_q-1$ capture only the residual information.
Several approaches can be adopted to address this issue, one of which is discussed explicitly in Appendix~\ref{app:helmert}.

\subsection{Case study: pairwise gradient of the inverse Laplacian}\label{sec:gradient}

In the previous sections, we have introduced a flexible strategy to incorporate additional information into the standard 2-point analysis through an auxiliary field derived from a generic transformation of the density field.
As a proof of concept, for the rest of this work we focus on a specific choice of auxiliary field, coupled with a particular pooling kernel.

We define the auxiliary field $\ve{\psi}$ as the gradient of the inverse Laplacian of the galaxy density contrast $\delta$ (with smoothing applied, see below), 
\begin{equation}\label{eq:psi}
    \psi_x(\ve{x}) = \int \frac{dk^3}{2\pi} e^{i \ve{k} \cdot \ve{x}} \frac{i k_x}{k^2} \delta(\ve{k}) \ ,
\end{equation}
and similarly for the other Cartesian components $\psi_y$ and  $\psi_z$, where it is implied that any quantity written as a function of the wavenumber 
$k$ is evaluated in Fourier space. 
This is the solution of the equation 
\begin{equation}\label{eq:psi_diff}
    \nabla \cdot \ve{\psi} = - \delta \ ,
\end{equation}
for fields that can be derived from a scalar potential, $\ve{\psi} = \nabla \phi$.
The latent variable $\lambda$ is defined through the following pooling kernel, 
\begin{equation}\label{eq:A}
    K_p = \left(\ve{\psi}_1 - \ve{\psi}_2\right) \cdot \frac{\ve{s}}{s} \ , 
\end{equation}
which represents the pairwise gradient along the separation vector.

The underlying idea behind this choice of auxiliary field and pooling function is to isolate infalling and outflowing pairs.
In fact, with the above definition, $\ve{\psi}$ can be seen as the solution of the usual Zel'dovich reconstruction equation,
$\nabla \cdot \ve{\psi} + \frac{f}{b} \frac{\partial}{\partial z}\psi_z = - \frac{\delta}{b}$, 
or of its more general formulation with a varying line of sight (see e.g. \cite{nusser1994, padmanabhan2012}),
for a galaxy population with $b=1$ and $f=0$, i.e. with no bias and redshift space distortions.
As such, it represents a proxy for the velocity field, modulo its overall amplitude.
Since two pairs with identical separation and distance from the observer but different angular positions on sky will be subjected to identical AP but different RSD effects, depending on their pairwise velocity, we expect this simple auxiliary field to help disentangling the two phenomena.

This picture is intentionally schematic, meant solely to provide physical insight and to explain the rationale behind the chosen latent variable.
In practice, we never assume that $\lambda$ represents a pairwise velocity or the true pairwise Zel’dovich displacement; rather, we treat it as a generic transformation of the galaxy field.
Indeed, when galaxy velocities can be measured directly (e.g. \cite{tully2023, saulder2023}), they contain additional information beyond that encoded in the galaxy field itself, and it may therefore be advantageous to analyse them separately.
As discussed earlier, we restrict ourselves to the case in which the auxiliary field is deterministically derived from the galaxy field and, by construction, does not introduce new information, but instead helps compress it into two-point statistics and facilitates its extraction.

One important point is that, because spectroscopic galaxy fields are measured in redshift space and are affected by the AP effect, these distortions inevitably propagate into any transformation of the field, thereby impacting its constraining power and making theoretical modelling more challenging.
The specific auxiliary field considered here makes no exception.
In particular, we expect its effectiveness in disentangling the two types of distortions to be weakened compared to the idealized picture introduced above.

As anticipated, a Gaussian smoothing is applied to the galaxy distribution.
Although not formally required, this procedure helps suppressing small-scale fluctuations, which are typically dominated by galaxy formation physics, nonlinear evolution, and observational effects.
Removing these contributions makes the interpretation in terms of galaxy infall clearer and simplifies the overall analysis.
The smoothing scale is left as a free parameter, but in realistic scenarios we expect values of order $10 h^{-1}\mathrm{Mpc}$ to be suitable.

The information encoded in the choice of auxiliary field and latent variable just described clearly overlaps with that used in standard reconstruction algorithms, which explicitly employ the Zel’dovich displacement field.
Indeed, one could extend Eq.~\ref{eq:psi_diff} to incorporate bias and growth, thereby mimicking the usual reconstruction equation mentioned above.
We briefly discuss this possibility in Section~\ref{sec:results}. For this proof of concept, however, we choose to keep the model as simple as possible, avoiding the introduction of cosmology-dependent parameters, and thereby emphasizing the spirit of the augmented correlation as a framework based on generic transformations of the field.

For similar reasons, we also expect some overlap with void-based analyses, which, broadly speaking, rely on identifying regions where galaxies flow outward toward the boundaries of voids. Density-split methods and density-marked correlation functions likewise exploit related information, as overdense (underdense) regions are expected, on average, to host more infalling (outflowing) galaxies.

That said, it is worth emphasizing that the cosmic structures probed by the augmented correlation can be more general in the following sense.
A standard two-point function partitions the pair space $\mathbb{R}^3 \times \mathbb{R}^3$ into smaller hypervolumes (bins) defined by pair separations and quantifies
how densely these bins are populated.
The augmented correlation builds on the same principle, but introduces an additional level of subdivision of the hypervolumes into smaller structures, determined by the distribution of the latent variable $\lambda$ in the six-dimensional space. In the quantile formulation, each region occupies an equal fraction $1/N_q$ of the volume of the bin.
Depending on the choice of the latent variable, these structures may be inherently pairwise in nature and, as such, may not correspond to well defined three-dimensional objects such as clusters, filaments, voids, or galaxy populations selected by environmental properties like density.
The specific definition of $\lambda$ discussed here, is indeed intrinsically pairwise: a given galaxy may be approaching one neighbour while receding from another (see Section~\ref{sec:results} for further discussions).

In general, we regard as non-fundamentally pairwise those latent variables that can be constructed from pooling kernels that do not couple $\psi(\ve{x}_1)$ and $\psi(\ve{x}_2)$.
For example density splits fit into this category.
We can write (some flavour of) density splits in the language of the augmented correlation by defining a two-dimensional kernel 
$K_p : V \times V \rightarrow \mathbb{R}^2$, where the vector space $V$ now refers to a scalar auxiliary field $\psi=\delta$, with $\delta$ denoting the density smoothed over a given scale.
Explicitly, the kernel takes the trivial form
$(K_{p1}, K_{p2}) = (\psi_1, \psi_2)$ or, to enforce symmetry under the  $\ve{x}_1 \leftrightarrow \ve{x}_2$ exchange, $(K_{p1}, K_{p2}) = [\min(\psi_1, \psi_2),\max(\psi_1, \psi_2)]$.
Although this version of density splits is not formally equivalent to that proposed by \cite{paillas2021}, which is constructed from cross correlations between galaxies and random tracers classified by local density, it effectively probes similar information.

Marked correlation functions, on the other hand, can be described in terms of one-dimensional kernels, most commonly of the form $K_p = \psi_1 \psi_2$, possibly with some suitable normalisation factor, which we omit for simplicity.
In the simplest case, one adopts $\psi = \delta$, as in density-split approaches, although other choices are equally viable.
For instance, the pairwise gradient considered in this work could itself serve as a mark.
Broadly speaking, the marked correlation function can be viewed as a compressed representation of the augmented correlation.
This can be seen by extending Eq. \ref{eq:xi_mean} to include higher $\lambda$ moments,
$1 + \xi_A^{(j)}(s,\mu) = \frac{1}{N_q}\sum_{i=0}^{N_q-1} {\lambda_i}^j \left[1 + \mc{Q}_i(s,\mu)\right]$,
where $\lambda_i$ denotes the mean value of the latent variable in the $i$-th quantile, possibly including a convenient normalization.
In this formulation for $j=0$ we recover the original equation, $\xi_A^{(0)} = \xi$, while $\xi_A^{(1)}$ captures the information contained in the marked correlation function (for sufficiently large $N_q$).
Whether compressing to a single $\lambda$ moment leads to information loss depends on the chosen latent variable.

\section{Fisher formalism}\label{sec:fisher}

Fisher information analysis \cite{Fisher1935, tegmark1997a, tegmark1997b} is a widely used statistical framework to forecast the constraining power of a given observable data vector $\mathbf{d}$ on a set of model parameters $\boldsymbol{\theta}$.
The basic idea is to quantify how sensitively the observable responds to small variations in the underlying parameters by looking at the derivatives of the log-likelihood 
$\partial \log \mathcal{L}(\mathbf{d}|\boldsymbol{\theta}) / \partial \boldsymbol{\theta}$, usually  referred to as the scores.
These are combined to construct the Fisher information matrix,
\begin{equation}
F_{ij}(\boldsymbol{\theta}) =
\left\langle
\frac{\partial \log \mathcal{L}(\mathbf{d}|\boldsymbol{\theta})}{\partial \theta_i}
\frac{\partial \log \mathcal{L}(\mathbf{d}|\boldsymbol{\theta})}{\partial \theta_j}
\right\rangle \ ,
\end{equation}
where the expectation value is taken over realizations of the data.

Under the assumption that the likelihood is approximately Gaussian near its maximum, the Fisher matrix
takes the simpler form\footnote{Following \cite{carron2013}, here we adopt a conservative approach by neglecting the term $\frac{1}{2} \mathrm{Tr} \left[
\mathbf{C}^{-1}
\frac{\partial \mathbf{C}}{\partial \theta_i}
\mathbf{C}^{-1}
\frac{\partial \mathbf{C}}{\partial \theta_j}
\right]$, which would formally appear in Eq.~\ref{eq:Fisher_Gauss} as an additional term to be summed to the expression.}
\begin{equation}\label{eq:Fisher_Gauss}
F_{ij}(\boldsymbol{\theta}) =
\frac{\partial \mathbf{d}}{\partial \theta_i}^{\top}
\mathbf{C}^{-1}
\frac{\partial \mathbf{d}}{\partial \theta_j} \ ,
\end{equation}
where $\mathbf{C}$ is the covariance matrix of the data vector $\mathbf{d}$.
This makes explicit that higher Fisher information arise both from strong parameter dependence and from low noise levels.
The resulting parameter constraints can be inferred via the Cramér–Rao bound, which directly maps the Fisher matrix into lower limits on the uncertainties,
\begin{equation}
\sigma_{\theta_i} \ge \sqrt{(F^{-1})_{ii}} , .
\end{equation}

While they rely on simplifying assumptions, most notably Gaussianity and the validity of a local expansion, Fisher forecasts provide a useful first step in evaluating the information content of new statistics and guiding their design.
In galaxy clustering analyses, $\boldsymbol{\theta}$ generally represents the set of parameters of a given cosmological model, often supplemented by additional parameters describing the connection between galaxies and the underlying dark matter field, such as galaxy bias and Halo Occupation Distribution (HOD, \cite{berlind2002}) parameters.

In this work $\boldsymbol{\theta}$ is the set of the $\nu\Lambda$CDM cosmological parameters and the observables $\mathbf{d}$ correspond the correlation functions, both traditional and augmented. 
Our goal is to quantify their relative performance, rather than their absolute constraining power, which is inherently survey-dependent.  
As detailed in the next section, we estimate both derivatives and the covariance matrix numerically using a large ensemble of dark matter halo simulations.
Although this setup does not fully capture the complexity of spectroscopic survey data, it is sufficiently realistic for the proof-of-concept analysis presented here.

\subsection{Simulations}

To quantify the information content of the statistics under consideration, we make use of the Quijote simulations \cite{villaescusa-navarro2020}, a large suite of cosmological N-body simulations specifically designed to probe the dependence of large-scale structure observables on cosmological parameters.
They consist of thousands of realizations that systematically vary key parameters such as matter and baryon density, dimensionless Hubble constant, spectral index, amplitude of density fluctuations and neutrino mass ($\Omega_m$, $\Omega_b$, $h$, $n_s$, $\sigma_8$, $M_\nu$), while keeping the remaining parameters fixed.

We make use of the fiducial-resolution subset of the suite, where each simulation evolves $512^3$ dark matter particles (plus $512^3$ particles for simulations with massive neutrinos) in a periodic box of side length $1 h^{-1}$Gpc, corresponding to a particle mass of $\sim 10^{11} h^{-1}M_\odot$.
The simulations are initialized at high redshift using second-order Lagrangian perturbation theory (first order for simulations with massive neutrinos) and evolved with the TreePM code \texttt{Gadget-III} \cite{springel2005}.
Dark matter halos are identified using a Friends-of-Friends (FoF) algorithm~\cite{davis1985} with a linking length of $b = 0.2$. The public halo catalogs include structures composed of at least 20 particles, corresponding to a minimum mass of $\sim 1.3 \times 10^{13} h^{-1}M_\odot$.

For each cosmological parameter, paired sets of simulations with positive and negative variations around the fiducial model are provided (see Table \ref{tab:quijote_params}), allowing numerical derivatives to be computed via centered differences.
In addition, the availability of a large number of independent realizations with the same fiducial cosmology allows for accurate estimation of covariance matrices.
\begin{table}[t]
\centering
\begin{tabular}{lcc}
\hline
Parameter & Fiducial value & Variation \\ 
\hline
$\Omega_m$     & 0.3175 & $\pm 0.01$ \\
$\Omega_b$     & 0.049  & $\pm 0.002$ \\
$h$            & 0.6711 & $\pm 0.02$ \\
$n_s$          & 0.9624 & $\pm 0.02$ \\
$\sigma_8$     & 0.834  & $\pm 0.015$ \\
$M_\nu \, [\mathrm{eV}]$ & 0.0    & 0.1, 0.2, 0.4 \\
\hline
\end{tabular}
\caption{Fiducial cosmological parameters of the Quijote simulations and corresponding variations used to compute derivatives.}
\label{tab:quijote_params}
\end{table}
With obvious notation, for $\Omega_m$, $\Omega_b$, $h$, $n_s$ and $\sigma_8$ the derivatives can be evaluated as
\begin{equation}
\frac{\partial \mathbf{d}}{\partial \theta_i}
\simeq
\frac{\mathbf{d}(\theta_i + \Delta \theta_i) - \mathbf{d}(\theta_i - \Delta \theta_i)}
{2\,\Delta \theta_i} \, ,
\label{eq:derivative_symmetric}
\end{equation}
whereas for the $M_\nu$ derivative (neutrino mass cannot be negative) one can use
\begin{equation}
\frac{\partial \mathbf{d}}{\partial M_\nu}
\simeq
\frac{
\mathbf{d}(4\Delta M_\nu)
- 12\,\mathbf{d}(2\Delta M_\nu)
+ 32\,\mathbf{d}(\Delta M_\nu)
- 21\,\mathbf{d}(M_\nu = 0)
}
{12\,\Delta M_\nu} \ . 
\label{eq:derivative_neutrino}
\end{equation}
One subtlety in constructing the data vectors for the latter equation is that, as mentioned above, the massive neutrino and fiducial simulations adopt different orders of Lagrangian perturbation theory for their initial conditions.
This is addressed in the Quijote suite by including a set of fiducial simulations with Zel’dovich initial conditions, which can be used to obtain a consistent $M_\nu=0$ data vector.

In this work, we focus on the $z = 1$ halo snapshot, which lies well within the redshift range probed by ongoing surveys such as DESI and Euclid.
Compared to similar analyses performed at lower redshift, we expect this choice to somewhat penalize the performance of the augmented correlation, since the non-Gaussian component of the field grows as $z \rightarrow 0$ (and, for a fixed mass threshold, the halo bias decreases).
Our goal here is simply to provide a proof of concept rather than to deliver a precise forecast of the expected errors for a given dataset or a consistent comparison with other alternative statistics, which would require a more realistic and sophisticated analysis.

\section{Results}\label{sec:results}

All measurements presented here, except those corresponding to the fiducial cosmology, are obtained by averaging over the 500 publicly available realizations of the Quijote simulations, each ``observed'' using the three Cartesian axes as independent lines of sight, for a total of $N_{\mathrm{real}} = 1500$.
For the fiducial cosmology, which is used to estimate the covariance matrices, we instead employ $N_{\mathrm{fid}} = 5000$ realisations.
The corresponding convergence properties are discussed in Appendix~\ref{app:convergence}.

We include all halos in the catalogs, yielding approximately $\sim 2 \times 10^5$ objects per realization, corresponding to a number density of $2 \times 10^{-4} h^{3}\mathrm{Mpc}^{-3}$.
For simplicity, we will hereafter occasionally refer to these halos as galaxies.
Volumes are sampled using a random catalog with number density three times higher than that of the data.
This choice is primarily motivated by the need to keep the computational cost manageable given the large number of repeated measurements.
While a higher density of randoms may be preferable, particularly in analyses of real data, it would not significantly change the results presented here.

To balance large-scale noise against small-scale information, we bin the pair separation $s$ into 50 logarithmic bins over the range $[s_{\min}, s_{\max}] = [1, 150] h^{-1}\mathrm{Mpc}$, although in practice we restrict the analysis to roughly half of these, corresponding to $s \gtrsim 10 h^{-1}\mathrm{Mpc}$.
The angular dependence is sampled using 200 linearly spaced bins over $\mu \in [-1,1]$.
Following standard practice, we integrate over $\mu$ to obtain the Legendre multipoles of the correlation function,
\begin{equation}
\xi_\ell(s) = \frac{2\ell + 1}{2} \int_{-1}^{1} d\mu \ \xi(s,\mu) \ \mc{L}_\ell(\mu) \ ,
\end{equation}
where $\mc{L}_\ell$ is the $\ell$-th Legendre polynomial.
We apply the same compression to $\xi_A$.
For the latent variable $\lambda$, we adopt 16,200 linear bins over the interval $[\lambda_{\min}, \lambda_{\max}] = [-50, 50] h^{-1}\mathrm{Mpc}$.
This fine binning allows for bins to be combined a posteriori, after computing the pair counts, in order to obtain the desired number of quantiles.

The latent variable $\lambda$ is constructed via the auxiliary field $\ve{\psi}$ and the pooling kernel $K_p$ introduced in Section~\ref{sec:gradient}.
We use Fast Fourier Transforms (FFTs) to perform both forward and inverse operations on a $512^3$ grid, corresponding to a spatial resolution of $\sim 2 h^{-1}\mathrm{Mpc}$.
To construct $\delta$, halo positions are painted onto the grid using a Cloud-In-Cell (CIC) scheme with unit weights.
Once in Fourier space, we apply a single deinterlacing step \cite{sefusatti2016} and correct for grid assignment effects via deconvolution \cite{jing2005}.
A Gaussian smoothing with scale $\sigma_{\mathrm{smooth}} = 10 h^{-1}\mathrm{Mpc}$ is then applied, followed by the $i k_n / k^2$ factors.
After transforming back to configuration space, the three Cartesian components of $\ve{\psi}$ are interpolated at the particle positions (halos and randoms) using trilinear interpolation.

We always enforce periodical conditions and assume fixed line of sight, parallel to one of the axis of the simulation box.
Redshift space coordinates $\ve{r}_s$ are obtained via 
the usual mapping 
\begin{equation}
\ve{r}_s = \ve{r} + \frac{v_{\parallel}(\ve{r})}{aH} \, \hat{n} \ ,
\end{equation}
where $\ve{r}$ denotes the real-space position, $\hat{n}$ is a unit vector along the chosen line of sight, $v_\parallel$ is the corresponding component of the peculiar velocity, and $aH$ is the conformal Hubble parameter.
To model the AP effect, we rescale the simulation box along and perpendicular to the line of sight according to
\begin{equation}
\alpha_\parallel = \frac{H^{\mathrm{fid}}(z)}{H(z)} \ , \qquad
\alpha_\perp = \frac{D_A(z)}{D_A^{\mathrm{fid}}(z)} \ ,
\end{equation}
where $D_A(z)$ is the angular diameter distance, and quantities labeled with the superscript “fid” are evaluated in the fiducial cosmology.
In practice, rather than explicitly modifying the halo positions, we apply these scalings by rescaling pair separations during the pair-counting stage, $(s_\parallel, s_\perp) = (\alpha_\parallel 
s_\perp^{\mathrm{fid}}, \alpha_\perp \, s_\perp^{\mathrm{fid}})$,
and by consistently adjusting the wavelengths of the Fourier modes when computing the auxiliary field $\ve{\psi}$.
This approach is formally equivalent to a full rescaling, provided that the random catalogue is generated to be uniform in the rescaled volume rather than in the original $1 h^{-1}\mathrm{Gpc}$ cubic box.

Regardless of whether one performs a full rescaling or adopts the procedure described above, the grid used for the FFTs (or, equivalently, the associated Fourier modes) becomes effectively adaptive, and remains strictly regular only in the fiducial cosmology. Whether it is preferable, in a realistic application, such as when constructing an emulator, to instead keep the cell size fixed is an open question.
In any case, we do not expect this choice to significantly affect the conclusions of the proof-of-concept presented here.

\begin{figure}[htbp]
\centering
\includegraphics[width=0.28\textwidth]{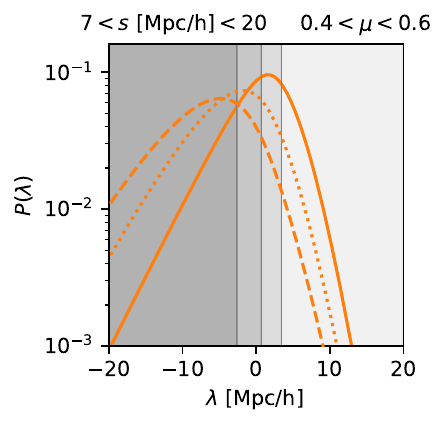}
\includegraphics[width=0.285\textwidth]{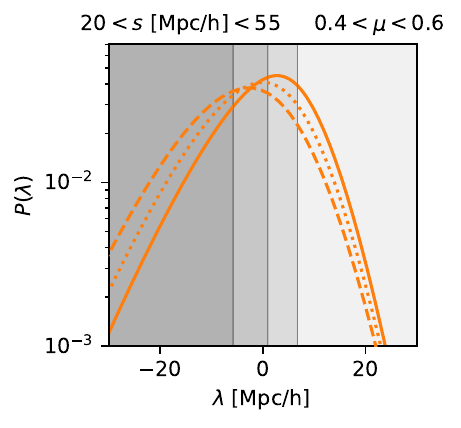}
\includegraphics[width=0.41\textwidth]{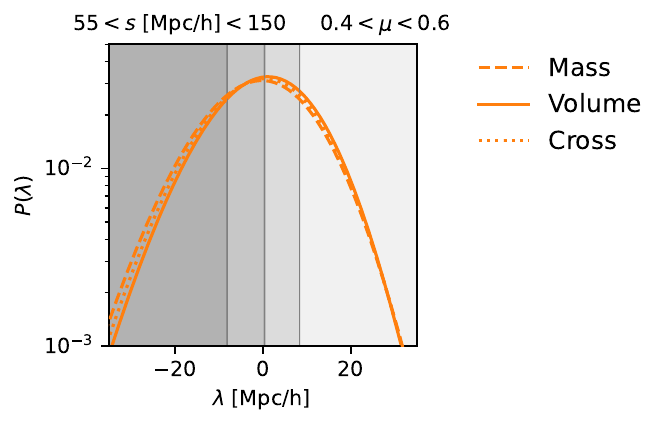}
\caption{Distributions of the latent variable $\lambda$, averaged over $N_{\mathrm{fid}} = 5000$ halo catalogues from the Quijote simulations at the fiducial cosmology ($z=1$). The three curves correspond to the components entering the construction of the augmented correlation: the mass-weighted, volume-weighted, and cross distributions, shown as dashed, solid, and dotted lines, respectively. The panels display different pair separations, increasing from left to right as indicated in the figure. The gray shaded regions mark the quantiles of the volume-weighted distribution for the case $N_q=4$. Note that the axis scales differ between panels.}
\label{fig:lambda_dist}
\end{figure}

Figure~\ref{fig:lambda_dist} shows the distributions of the latent variable $\lambda$ at different pair separations averaged over the $N_\mathrm{fid} = 5000$
fiducial-cosmology simulations.
The three curves correspond to the three fundamental building blocks entering the definition of the augmented correlation (see Section~\ref{sec:model} and Appendix~\ref{app:estimator}): the mass-weighted (dashed), volume-weighted (solid), and cross distributions.
Their overall shape is consistent with the interpretation of $\lambda$ as a proxy for pair infall, being approximately Gaussian near the core but developing exponential tails.
By construction, negative and positive values of $\lambda$ correspond to infall and outflow, respectively.
As the separation decreases, moving from the right to the left panel, the exponential component becomes increasingly prominent and the distributions progressively more skewed, as expected from the growing importance of nonlinear motions on small scales.

All three panels display the same angular bin, $\mu \in [0.4,0.6]$, since varying $\mu$ produces nearly identical distributions.
This is a direct consequence of the specific pooling kernel $K_p$ adopted in this work, which projects the three-dimensional infall along the pair separation vector.
With this choice, in the absence of RSD and AP effects, the distributions would be strictly independent of $\mu$.
By contrast, if the infall were projected along the line of sight, the odd central moments of the distributions would acquire an intrinsic $\mu$-dependence (see e.g. \cite{scoccimarro2004, bianchi2015a, bianchi2016}). 
Nevertheless, this more complex scenario would not require any modification of the tools introduced in this work, as the quantiles are computed separately in each $(s,\mu)$ bin. 

The $\lambda$-quantiles of the volume-weighted distributions are indicated by the gray shaded regions for the case $N_q=4$.
Although the logarithmic scale adopted on the $y$-axis is useful for highlighting the exponential tails of the distributions, it makes the quantiles less immediately recognizable as equal-probability regions.
Nevertheless, the figure clearly illustrates how the quantiles evolve with separation (note the different axis scales in the various panels).
As discussed in Section~\ref{sec:quantiles}, the quantiles of the volume-weighted distributions define the $\lambda$-bins of the mass-weighted and cross distributions in an adaptive, separation-dependent manner.
Qualitatively, the fact that, at small separations, the mass-weighted distribution is shifted toward more negative $\lambda$ values relative to the volume-weighted one implies that, at those separations, the augmented correlation is expected to exhibit a larger amplitude in the first quantile, progressively decreasing toward higher quantiles, as corroborated by the results shown below.

Figure~\ref{fig:xi_ebars} illustrates the monopole ($\ell = 0$), quadrupole ($\ell = 2$), and hexadecapole ($\ell = 4$) of the augmented correlation function (left panel, orange) for the $N_q = 4$ quantiles case, as described in Section~\ref{sec:quantiles}. These are compared to the corresponding multipoles of the standard correlation function (right panel, blue). All curves represent averages over the $N_\mathrm{fid} = 5000$ fiducial cosmology simulations, while the shaded regions indicate the $1\sigma$ uncertainties for a single simulation box.
\begin{figure}[htbp]
\centering
\includegraphics[width=0.625\textwidth]{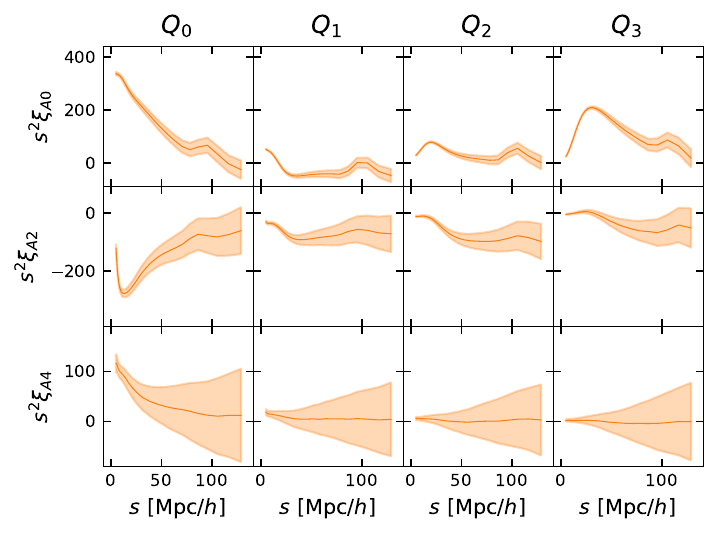} 
\includegraphics[width=0.365
\textwidth]{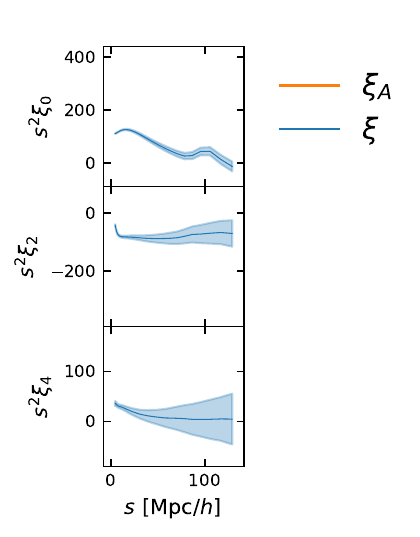}
\caption{Monopole ($\ell=0$; first row), quadrupole ($\ell=2$; second row), and hexadecapole ($\ell=4$; third row) of the augmented (orange; left block) and the standard correlation function (blue; right block), measured from $N_{\mathrm{fid}} = 5000$ halo catalogues of the Quijote simulations at the fiducial cosmology ($z=1$).
The curves represent the average over all realizations, while the shaded regions indicate the $1\sigma$ uncertainty for an individual realization.
For visualisation purposes, all the multipoles are multiplied by the separation squared.
The augmented correlation is compressed into $N_q=4$ quantiles of the latent variable, corresponding to the four columns in the left block, as indicated in the figure.
These columns probe different infall regimes, ranging from strong infall to outflow when moving from left to right, or equivalently from $\mc{Q}_0$ to $\mc{Q}_3$.}
\label{fig:xi_ebars}
\end{figure}

From the figure, it is clear that extending the correlation into the $\lambda$ latent space allows one to isolate distinct clustering features that are otherwise suppressed when averaging to obtain the standard correlation function.
Overall, we observe familiar behaviors: monopoles whose amplitude tends to vanish on large scales while highlighting the BAO peak, and negative quadrupoles arising from redshift-space distortions.
However, these features vary significantly across quantiles, sometimes changing sign, and, even at fixed quantile, exhibit a stronger scale dependence compared to the traditional correlation.

Adopting the interpretation of  
$\lambda$ introduced in
section~\ref{sec:gradient} and supported by Figure~\ref{fig:lambda_dist}, the different columns in the left panel of Figure~\ref{fig:xi_ebars} can be broadly viewed as tracing the clustering of the most infalling pairs on the left, followed by approximately static pairs in the middle, and finally the most outflowing pairs on the right.
One important point to bear in mind is that, as discussed in section~\ref{sec:gradient}, $\lambda$ is intrinsically pairwise; consequently, the structures probed by the different quantiles in the figure formally reside in a six-dimensional space and do not admit a straightforward one-to-one correspondence with, for example, clusters, voids, or other three-dimensional entities.
It is nevertheless reasonable to expect that some weaker correspondence may still hold; for instance, a galaxy residing in (or near) a cluster is likely to form predominantly infalling pairs on scales comparable to that of the cluster.
On larger scales, the same galaxy may instead form strongly outflowing pairs with galaxies moving toward their own local overdensities, effectively disconnected from that of the original galaxy.

Given how the augmented correlation is defined, what the figure is rigorously telling us is that, on small scales, regions of the six-dimensional pair space that are identified as infall regions are highly populated compared to outflow regions, as already suggested by Figure~\ref{fig:lambda_dist}, or, in other words, galaxy pairs with separation $\lesssim 30 h^{-1}\mathrm{Mpc}$ tend to reside in infall regions.
Not surprisingly, these regions show the most negative quadrupole, due to stronger RSD.
Conversely, on the same scales the quadrupole of the outflowing pairs becomes consistent with zero or even slightly positive for $\mc{Q}_3$. 

On larger scales, one can observe hints of the effect mentioned above, namely galaxies being attracted toward distinct minima of the gravitational potential. In this regime, the relative orientation of $\ve{\psi}$ is nearly random, and infalling and outflowing configurations become almost equally likely. What primarily determines the $\lambda$ quantiles is therefore the average magnitude of the auxiliary field, $|\ve{\psi}|$, which is smaller (larger) in low- (high-) density regions. In this sense, it is plausible that a significant fraction of the negative correlation observed in $\mc{Q}_1$ arises from pairs residing in voids, not necessarily the same one.
Once again, these considerations serve only as sanity checks; by design, what we are truly probing are pairwise void-like structures.

The hexadecapole is broadly consistent with the scenario described, showing strong RSD effects for the most infalling pairs.
However, it is also significantly noisier than the other multipoles, and for this reason we do not include it in the remainder of the analysis.

Interestingly, the position of the BAO peak appears to drift across the quantiles, shifting smoothly from smaller separations for infalling pairs to larger separations for outflowing ones. Similar behaviour has been reported in analyses investigating the environmental dependence of the BAO peak position (e.g. \cite{neyrinck2018}).
This trend is physically expected, as the $\lambda$ quantiles are effectively tracing the large-scale flows responsible for the broadening and displacement of the BAO feature, which reconstruction techniques are designed to partially reverse.
Whether this effect can be exploited to improve BAO measurements remains an open question that we leave for future investigation.

\begin{figure}[htbp]
\centering
\includegraphics[width=0.565\textwidth]{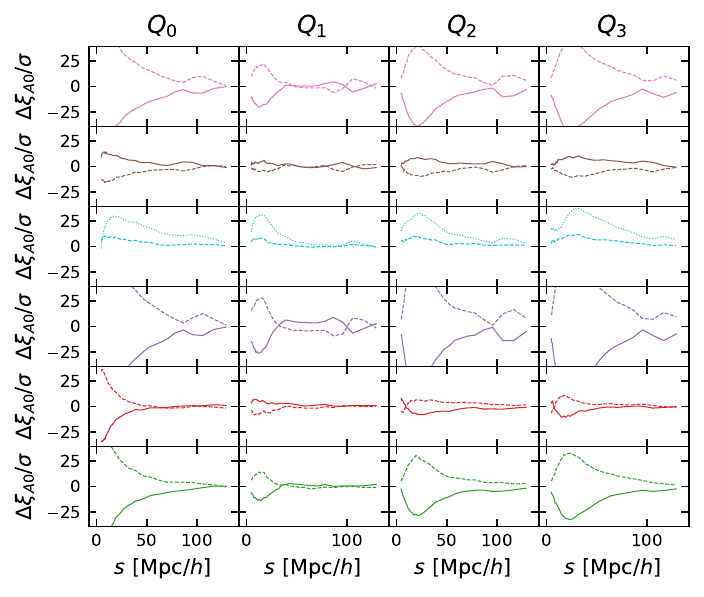} 
\includegraphics[width=0.42
\textwidth]{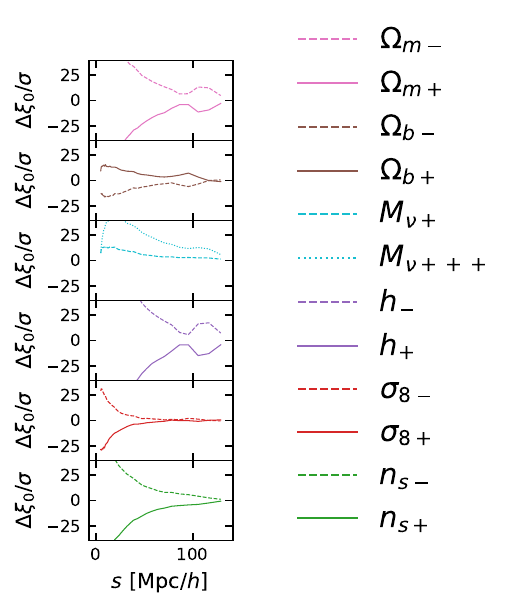}
\caption{Impact of variations in the cosmological parameters on the measured statistics.
The curves show the residuals $\Delta \xi = \xi_0 - \xi_{0,\mathrm{fid}}$ of the monopole for both the augmented (left block) and the standard correlation function (right block), obtained by averaging over the full set of $N_{\mathrm{real}} = 1500$ realizations.
The residuals are normalized by the corresponding standard deviation of the mean in order to quantify the signal relevant for the Fisher analysis relative to the intrinsic noise arising from the finite number of simulations.
For $\Omega_m$ (pink), $\Omega_b$ (brown), $h$ (purple), $\sigma_8$ (red), and $n_s$ (green), positive and negative parameter variations are shown with solid and dashed lines, respectively.
For $M_{\nu}$ (cyan), three positive variations are available; two are displayed here as dashed ($M_{\nu+}$) and dotted ($M_{\nu+++}$) lines, while the intermediate case ($M_{\nu++}$) lies between them and is omitted for clarity.
As in Figure~\ref{fig:xi_ebars}, the columns of the left block correspond to the $N_q=4$ quantiles of the latent variable.}
\label{fig:dev}
\end{figure}

As noted in Section~\ref{sec:model}, the ability of the $\lambda$ quantiles to isolate meaningful clustering features does not automatically imply that they encode usable cosmological information.
This ultimately depends on how they respond to variations of the underlying cosmology.
Figure~\ref{fig:dev} shows $\Delta \xi = \xi_0 - \xi_{0,\mathrm{fid}}$ (right panel), expressed in units of the standard deviation of the mean, for the different variations of the cosmological parameters, and analogously for $\xi_A$ (left panel).
These quantities are closely related to the derivatives entering the Fisher matrix, which can be recovered by dividing the separation between the two curves in each panel (with the exception of the neutrino case) by the parameter variation listed in Table~\ref{tab:quijote_params}.

The overall amplitude of the curves is largely set by binning choices and is therefore not particularly informative on its own.
More relevant are their relative amplitudes and the differing scale dependence of the $\xi_A$ quantiles compared to the standard correlation.
Overall, the augmented statistics displays a richer morphology, as expected if additional information is being captured. 
As an illustration, $\mc{Q}_1$ and $\mc{Q}_3$ show nearly opposite responses to changes in $\sigma_8$, with their contributions effectively cancelling when averaged to recover the standard correlation.
Moreover, $\mc{Q}_2$ retains sensitivity to these variations even on large scales, where the response of the standard correlation function is already close to zero.

\begin{figure}[htbp]
\centering
\includegraphics[width=0.49\textwidth]{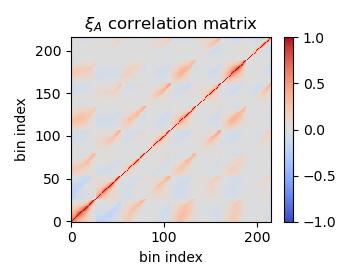}
\includegraphics[width=0.49\textwidth]{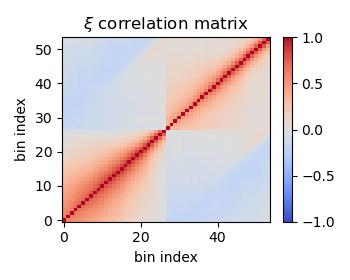} 
\caption{Covariance matrices, normalized by their diagonal elements, for the monopole and quadrupole of the augmented (left) and standard correlation function (right), measured from $N_{\mathrm{fid}} = 5000$ halo catalogues of the Quijote simulations at the fiducial cosmology ($z=1$).
The bin index of the $\xi$ matrix increases from small to large separations, with monopole bins listed before quadrupole bins.
In this ordering, the lower-left and upper-right diagonal blocks correspond to the monopole and quadrupole covariances, respectively, while the off-diagonal blocks encode their cross-covariances.
The separation range considered is $s \in [10,150] h^{-1}\mathrm{Mpc}$, corresponding to a total of 27 bins.
The same structure is adopted for the $\xi_A$ matrix, repeated $N_q=4$ times for the different quantiles, beginning with the highest-infall quantile $\mc{Q}_0$.}
\label{fig:covmat}
\end{figure}

The fact that the curves remain significantly above unity almost everywhere suggests that the derivatives are sufficiently well sampled and that the resulting Fisher information is genuine, but see Appendix~\ref{app:convergence} for a more detailed analysis. 
That said, any meaningful consideration on the information content requires proper weighting through the covariance matrix.
Figure~\ref{fig:covmat} shows the correlation matrix (i.e. the covariance matrix normalised by its diagonal elements) of quadrupole and monopole of both augmented and standard correlations.
The bin index of the $\xi$  matrix (right panel) runs from small to large separation with the monopole bins listed first.
In this arrangement, the lower-left and upper-right blocks (27 bins each) contains the monopole and quadrupole covariance, respectively, whereas the symmetric off-diagonal blocks capture the cross-covariances.
The same scheme applies to the $\xi_A$ matrix (left panel), but is repeated $N_q$ times, starting from the highest-infall quantile $\mathcal{Q}_0$.

As expected, the covariance of $\xi$ is dominated by its diagonal elements, with smooth off-diagonal contributions that decay for entries far from the diagonal, which encode the coupling between small and large scales, and negligible power in the monopole-quadrupole cross terms. 
The $\xi_A$ covariance displays a similar structure, replicated across the $N_q$ quantiles.
In addition, faint secondary diagonals, parallel to the main one, indicate a moderate coupling between analogous components across different quantiles.
These couplings weaken as the quantiles probe increasingly different $\lambda$ values, i.e., as one moves away from the diagonal.
In conclusion, the structure of the $\xi_A$ covariance is quite regular, which facilitates both its interpretation and, more importantly, its sampling and numerical inversion.
Moreover, the relatively modest correlation between the $\lambda$-quantiles suggests that the information they encode is not strongly redundant, allowing the corresponding signals shown in Figure~\ref{fig:dev} to combine constructively and enhance the overall Fisher information.

\begin{figure}[htbp]
\centering
\includegraphics[width=1\textwidth]
{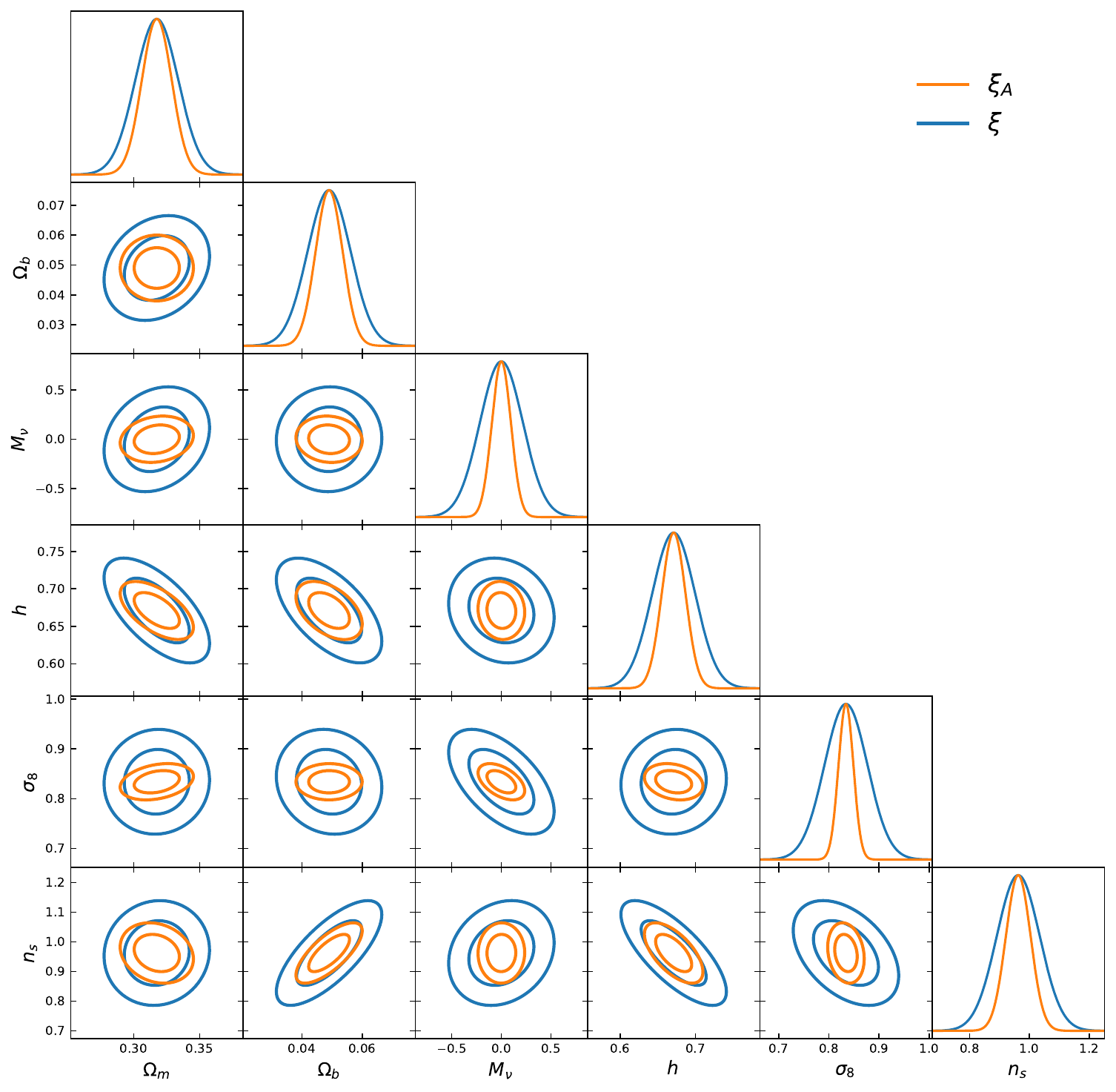}
\caption{Fisher forecast constraints from the Quijote halo catalogues at $z=1$ for the augmented (orange) and standard correlation function (blue) within the $\nu\Lambda\mathrm{CDM}$ cosmological model.
The off-diagonal panels show the two-dimensional parameter constraints with $1\sigma$ and $2\sigma$ confidence contours, while the diagonal panels display the corresponding one-dimensional marginalized distributions, normalized to unity at their maxima.
The forecasts are obtained by combining the monopole and quadrupole over the separation range $s \in [10,150],h^{-1}\mathrm{Mpc}$.
For the augmented correlation, $N_q=4$ quantiles of the latent variable are employed.}
\label{fig:triangle}
\end{figure}

Figure~\ref{fig:triangle} presents the Fisher forecasts for the six $\nu \Lambda \mathrm{CDM}$ cosmological parameters considered in this work, comparing the standard (blue) and augmented (orange) correlation functions in a conventional corner-plot format.
The lower panels show the two-dimensional constraints with 1 and 2$\sigma$ contours, while the diagonal panels display the corresponding one-dimensional distributions, normalized to unity at their peak.
To safely exclude small scales potentially affected by numerical artefacts in the simulations, for the figure we adopt a separation range of $[s_{\min}, s_{\max}] = [10, 150] h^{-1}\mathrm{Mpc}$.
The augmented correlation function yields tighter constraints for all cosmological parameters.

\begin{figure}[htbp]
\centering
\includegraphics[width=0.29\textwidth]{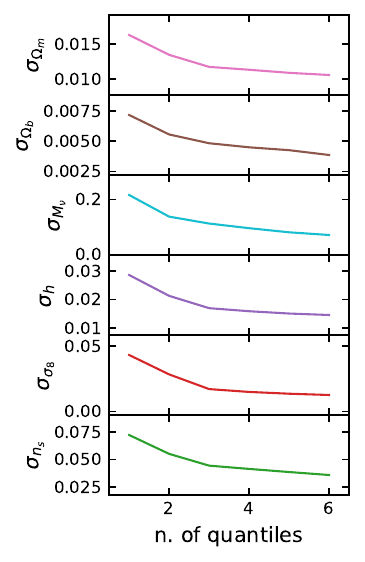}
\includegraphics[width=0.255\textwidth]{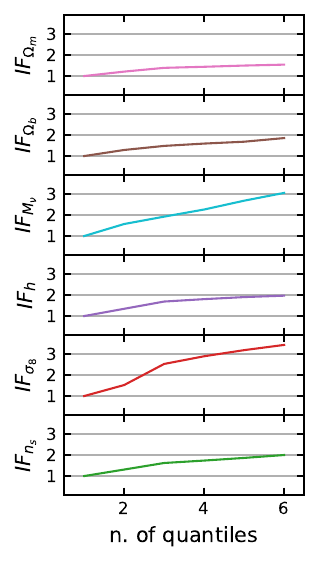}
\includegraphics[width=0.435\textwidth]{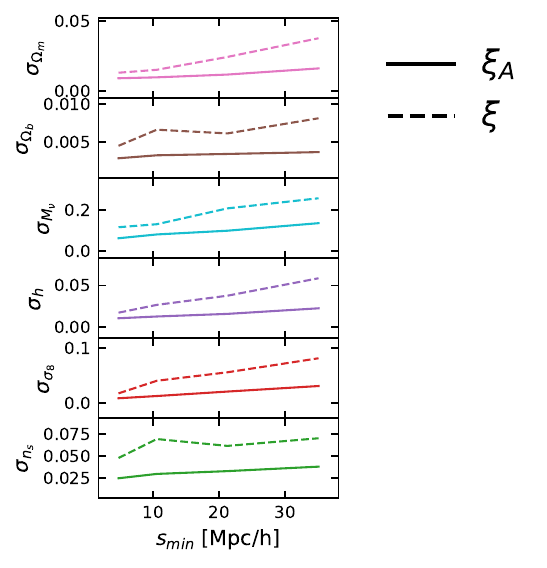}
\caption{Left column: $1\sigma$ Fisher constraints on the cosmological parameters obtained from the augmented correlation as a function of the number of latent-variable quantiles, $\sigma(N_q)$, for the fixed separation range $s \in [10,150] h^{-1}\mathrm{Mpc}$.
From top to bottom, the curves correspond to $\Omega_m$ (pink), $\Omega_b$ (brown), $M_{\nu}$ (cyan), $h$ (purple), $\sigma_8$ (red), and $n_s$ (green).
Central column: corresponding improvement factor, defined as $\mc{IF} = \sigma/\sigma_{\mathrm{ref}}$, where $\sigma_{\mathrm{ref}} = \sigma(N_q=1)$ denotes the Fisher constraints obtained from the standard correlation function.
Right column: $1\sigma$ Fisher constraints on the cosmological parameters obtained from both the augmented (solid) and standard (dashed) correlations as a function of the minimum separation scale, $\sigma(s_{\min})$, for a fixed number of quantiles $N_q=4$.}
\label{fig:Nq-scale}
\end{figure}

A more quantitative comparison is provided in Figure~\ref{fig:Nq-scale}, which shows the dependence of the constraints on the chosen configuration.
The left column of the figure illustrates how the constraints vary with the number of quantiles $N_q$, while the central column shows the corresponding improvement factor relative to the standard correlation function.
All results discussed in this section, except those explicitly shown the in the left and central panel of this figure, assume $N_q = 4$.
This choice is arbitrary and is adopted solely to simplify the discussion, without implying that $N_q = 4$ is in any sense optimal.
In an idealized case with noiseless data and unlimited computational resources, it is reasonable to expect that increasing the number of quantiles would allow more information to be captured.
This expectation is supported by the monotonic trend observed in the figure, where the inclusion of the $N_q = 5$ and $N_q = 6$ cases demonstrates a further improvement, reaching factors of $\sim 2-3$.

However, increasing $N_q$ also raises the number of data points.
This leads to noisier derivatives and less accurate covariance matrices, thereby increasing the risk of misinterpreting noise as signal or similar artefacts.
$N_q = 4$ represents a sufficiently conservative choice for the proof of concept presented here (see appendix~\ref{app:convergence}).
When dealing with real data, e.g. with an emulator, the optimal number of quantiles should be determined based on the specifics of the dataset under consideration, the number of available simulations, and, more broadly, will depend on the choice of latent variable.
   
The right column of Figure~\ref{fig:Nq-scale} illustrates how the constraints change as the minimum scale $s_{\min}$ is varied.
As expected, the constraints become weaker and weaker as the analysis is restricted to larger and larger scales; however, the information gain is preserved, or even increases in some cases.
This is counter intuitive, since the amount of non-Gaussian information should decrease with increasing separation, implying that the standard two-point correlation function alone should provide an adequate compression.
A similar trend was reported in \cite{paillas2023} for density-split statistics, where the authors also performed direct tests on purely Gaussian fields to gain further insight.
They argued that the very construction of density quantiles relies on small-scale information, which becomes intrinsically embedded in the statistic and is not lost even when the analysis is confined to large scales.

A similar line of reasoning applies to the analysis presented here, since the construction of the auxiliary field $\ve{\psi}$ incorporates information down to  (at least) the smoothing scale of $\sim 10 h^{-1}\mathrm{Mpc}$.
In addition, as discussed in Section~\ref{sec:model}, part of the information gain provided by the augmented statistic may originate from a more flexible or refined effective binning, rather than from genuinely non-Gaussian information.
This effect is plausibly amplified when the analysis is restricted to large scales, where the total number of bins is reduced, particularly under the logarithmic binning scheme employed in this work.
 
The choice of adopting the same separation binning for both statistics was made to avoid introducing additional sources of uncertainty associated with differing analysis setups.
Ultimately, however, determining what constitutes a truly fair comparison is a subtle issue. Independently of these configuration-dependent considerations, an overall information gain remains plausible for several reasons: beyond the implicit incorporation of small-scale information discussed above, the augmented correlation demonstrably succeeds in isolating distinct clustering regimes (Figures~\ref{fig:xi_ebars} and \ref{fig:dev}), while reconstruction-like transformations similar to those employed here to define the latent variable are already known to access higher-order information \cite{schmittfull2015, hikage2020}.
At the same time, the exact magnitude of the gains reported here is likely optimistic and may decrease once both statistics are fully optimized in terms of binning choices and realistic observational effects are taken into account.
A detailed investigation of these aspects lies beyond the scope of the present exploratory study, whose main goal is to introduce the framework rather than to provide a definitive quantitative assessment. More generally, the outcome is expected to depend sensitively on the specific choice of auxiliary field and latent variable.

The goal of the augmented correlation is to provide a flexible framework for incorporating arbitrary transformations of the galaxy field into clustering analyses.
To this end, we introduced the concept of an auxiliary field $\psi$, together with a pooling kernel $K_p$ that maps it into a latent variable $\lambda$, which is ultimately used to characterize the clustering. As a proof of concept, we have focused on the specific choice of $\psi$ and $K_p$ (and therefore $\lambda$) described in Section~\ref{sec:gradient}.
We defer to future work a more systematic exploration of the, effectively limitless, space of possible alternatives.
Nevertheless, we have explicitly examined a few cases that can be regarded as minor variations of the choices adopted here.
Due to computational limitations, these additional tests were carried out using a reduced set of simulations, amounting to approximately $10\%$ of those employed in the main analysis.
To ensure consistency, we adopt as a reference a correspondingly downgraded version of the main analysis, obtained with the same number of simulations.
However, this does not guarantee convergence, and the results should therefore be regarded as indicative rather than definitive.

Specifically, we have explored projecting $\ve{\psi}$ along the line of sight instead of along the separation vector, in order to better align with RSD effects.
This corresponds to $K_p = \left(\ve{\psi}_1 - \ve{\psi}_2\right) \cdot \hat{n}$, where $\hat{n}$ is the line-of-sight versor.
Within the intrinsic limitations of these tests discussed above, we do not observe appreciable changes in the constraints.
However, the Fisher contours (as well as the multipoles) exhibit slight differences in shape, hinting at the presence of residual information.
In practice, a natural way to capture this would be to extend the kernel from one to two dimensions by including the component perpendicular to the separation, which we have ignored here for simplicity.

We have also considered obtainig $\ve{\psi}$ from the standard BAO reconstruction equation, $\nabla \cdot \vec{\psi} + \frac{f}{b} \frac{\partial}{\partial z}\psi_z = - \frac{\delta}{b}$, which takes into account the fact that galaxies are biased tracers of dark matter and they are observed in redshift space.
Strictly speaking, this equation would require varying the growth rate across different cosmologies (and likewise the bias), but in real life the cosmology is precisely what we aim to constrain.
To avoid this circularity, we instead fix $f = f_{\mathrm{fid}}$ and $b = 3$, a realistic bias value, across all simulations.
Under this setup, we do not find clear indication of improved cosmological constraints.
This is one of the main reasons why, in our primary analysis, we adopt a simpler and cleaner definition of $\ve{\psi}$.
That said, it cannot be excluded that a formulation more closely resembling the actual displacement field such as the one outlined above, or a more sophisticated variant, may prove advantageous in practice, for example when working with real data or developing theoretical models.

We experimented with a Gaussian smoothing scale of  $\sigma_{\mathrm{smooth}} = 5h^{-1} \mathrm{Mpc}$.
While the multipole shapes change slightly, this has no significant impact on the Fisher contours.
We ultimately adopted $10h^{-1} \mathrm{Mpc}$ as a conservative choice, given the simulation number density and the grid resolution employed; however, strictly speaking, there is no strong evidence to favour this value.

\section{Summary and conclusions}\label{sec:conclusions}

Standard two-point clustering statistics, while powerful, compress the information encoded in spectroscopic surveys into a relatively restricted representation based solely on pair separations and orientations.
The structure generated by nonlinear evolution, galaxy bias, and redshift-space distortions is only partially captured by these observables.
In particular, standard two-point statistics are mainly sensitive to Gaussian features of the density field, while late-time evolution generates additional non-Gaussian information that can be averaged out in conventional analyses.
This motivates the development of alternative clustering statistics capable of reorganizing or conditioning the galaxy distribution in ways that preserve a richer description of the clustering signal.

Typical questions that arise when developing these approaches include: is there a transformation of the galaxy field that achieves optimal information compression? Which transformations best isolate particular physical effects? Which are more resilient to systematic uncertainties? And is there an optimal combination of different transformations that captures most of the information while remaining computationally efficient?

The space of possible transformations is, in principle, virtually unbounded.
To navigate it more systematically, we introduced a flexible framework that we refer to as the augmented correlation function that can be implemented using largely standard analysis pipelines, requiring only minor modifications to existing codes.
In this approach, the information encoded in an arbitrary transformation of the field is used to define a new variable, $\lambda$.
This “latent” variable effectively extends the traditional two-point correlation into one or more additional dimensions, $\xi(s,\mu) \rightarrow \xi_A(s,\mu,\lambda)$.
Its role is to capture and amplify clustering properties that are not solely determined by pairwise separations and would therefore be washed out when averaging to obtain the traditional two-point correlation function.
If these properties contain relevant cosmological information, the resulting constraining power is enhanced.

As a proof of concept we studied the case in which the latent variable is a measure of the infall of the pairs, constructed from the pairwise gradient of the inverse Laplacian of the galaxy density field.
We discussed how the augmented correlation can be organized into quantiles and used them to show that this choice of latent variable is indeed capable of isolating distinct clustering features, consistent with the expected behaviour of infall- and outflow-dominated regimes, such as enhanced or suppressed quadrupoles and systematic shifts of the BAO peak position. 

To assess the cosmological information content, we performed a standard Fisher analysis on halo catalogues at $z=1$ from the Quijote simulation, within a $\nu \Lambda\mathrm{CDM}$ cosmology.
Compared to the conventional two-point correlation function, the augmented statistic delivers systematically tighter constraints for all cosmological parameters considered, with improvements reaching factors of $\sim 2 – 3$ for the most constraining configurations explored in this study.

It is nevertheless important to emphasize that the analysis presented here is subject to the well known limitations affecting this type of exploratory studies.
Beyond the theoretical approximations inherent to the Fisher formalism, the estimation of the Fisher information itself is affected by a number of numerical limitations, which can be broadly divided into two categories.
First, although standard convergence tests have been performed, the finite number of simulations, the noise affecting each realization, and the finite-difference procedure used to estimate the derivatives inevitably introduce sampling uncertainties in the results.
Second, the simulations do not provide fully realistic representations of galaxy surveys, owing to their simplified geometry, the use of halos as tracers, and the absence of observational systematics.
Moreover, the constraining power depends on separation binning and number of $\lambda$-quantiles adopted, choices that have not been optimised in this work.

For these reasons, the gains reported here should be viewed as indicative of the promise of the method, while their exact magnitude should not be straightforwardly extended to realistic survey conditions.
We are currently developing an emulator for the augmented statistics that will enable a more direct assessment of the information carried by the latent variable considered in this work, beyond the intrinsic limitations of Fisher forecasts.
This will also allow us to study its response to realistic survey windows and selection effects, with the ultimate goal of performing cosmological inference on actual spectroscopic data.

In parallel, we plan to investigate alternative choices of latent variables, including a more systematic exploration of the variations briefly discussed in Section~\ref{sec:conclusions}, as well as entirely new directions.
Natural extensions include using the smoothed density field to establish a more direct connection with density-split and marked correlation statistics, as outlined in Section~\ref{sec:model}, or considering latent variables constructed from higher-order derivatives of the inverse Laplacian and, more generally, from topological and morphological descriptors, including void finders.
In theory, one could even envision making the transformation itself learnable.
Finally, because the augmented correlation is built from relatively simple and well-defined mathematical operations, deriving an analytic model for it is conceivable.
In practice, however, this task is complicated by the fact that the transformation can only be applied to the observed redshift-space galaxy field, including all associated observational effects, making this route more challenging.

\appendix

\section{Derivation of the estimator}\label{app:estimator}

Below we briefly outline the derivation of the estimator for the augmented correlation.
By substituting $\delta = \rho/\langle \rho \rangle - 1$, Eq.~\ref{eq:def_general} becomes
\begin{equation}
\xi_A = \frac{\left\langle \frac{\rho_1 \rho_2}{\langle\rho_1\rangle \langle\rho_2\rangle} \delta_D \left(\lambda - K_{p12}\right) \right\rangle
- \left\langle \frac{\rho_1}{\langle\rho_1\rangle} \delta_D \left(\lambda - K_{p12} \right) \right\rangle
- \left\langle \frac{\rho_2}{\langle\rho_2\rangle} \delta_D \left(\lambda - K_{p12} \right) \right\rangle
+ \left\langle \delta_D \left(\lambda - K_{p12} \right) \right\rangle}
{\left\langle \delta_D \left(\lambda - K_{p12} \right) \right\rangle} \ ,
\end{equation}
where, for compactness, we introduced the notation $\rho_i = \rho(\ve{x}_i)$, and similarly for $K_p$.
To gain intuition, let us focus on the first term in the numerator. Multiplying and dividing by $\langle \rho_1 \rho_2 \rangle$, it can be rewritten in terms of
$\left\langle \rho_1 \rho_2 , \delta_D(\lambda-K_{p12}) \right\rangle
/ \left\langle \rho_1 \rho_2 \right\rangle$, 
which corresponds to the normalized mass-weighted distribution of $\lambda$ for pairs located at $\ve{x}_1$ and $\ve{x}_2$.
In practice, this quantity can be estimated as
$DD(\ve{x}_1,\ve{x}_2,\lambda) / [DD(\ve{x}_1,\ve{x}_2) \Delta_\lambda]$, 
where $\Delta_\lambda$ denotes the width of a sufficiently small $\lambda$-bin.
An analogous interpretation applies to the remaining terms in the equation, with the last term and the two central terms generating the volume-weighted and cross distributions, respectively.

Combining all these contributions, the factors $DD(\ve{x}_1,\ve{x}_2)$, $DR(\ve{x}_1,\ve{x}_2)$, $RR(\ve{x}_1,\ve{x}_2)$, and $\Delta_\lambda$ cancel out, yielding the familiar expression
\begin{equation}
\hat{\xi}_A(s, \mu, \lambda) =
\frac{DD(s, \mu, \lambda) - 2DR(s, \mu, \lambda) + RR(s, \mu, \lambda)}
{RR(s, \mu, \lambda)} \ ,
\end{equation}
where we have switched to the standard $(s,\mu)$ coordinates, implicitly assuming ergodicity.

\section{Alternative basis for the augmented correlation}\label{app:helmert}

There are several possible ways to combine the latent-variable quantiles such that the standard correlation function, $\xi$, appears explicitly as one of the $N_q$ resulting statistics, while the remaining $N_q-1$ encode only the additional information.
Here we consider a simple option, often referred to as the Helmert basis,
\begin{align}
\ve{h}_0 &= \frac{1}{N_q} (1, \dots, 1), \\
\ve{h}_k &= \frac{1}{\sqrt{N_q (k+1)k}} 
(\underbrace{1, \dots, 1}_{k}, -k, 0, \dots, 0), 
\quad k=1,\dots,N_q-1.
\end{align}
This basis is orthogonal (and becomes orthonormal upon multiplication by $\sqrt{N_q}$) and is uniquely defined, up to the ordering of its vectors.
Its main properties follow directly from the definition: the first vector returns the average of any vector projected onto it, while the remaining vectors act as response functions, since their elements sum to zero by construction.

More explicitly, if we replace the original $\mc{Q}$ statistics with $N_q$ new quantities defined as
\begin{equation}
    \mc{S}_i = \sum_{j=i}^{N_q-1} M_{ij}\mc{Q}_j \ ,
\end{equation}
where $M$ is the matrix constructed from the basis vectors introduced above, we find by construction that $\mc{S}_0 = \xi(s,\mu)$.
To understand the meaning of the $\mc{S}_{i>0}$ functions, it is useful to consider the limit in which the auxiliary $\ve{\psi}$ field is spatially uncorrelated with the density field.
In this case, all the $\mc{Q}$ functions reduce to copies of the standard two-point correlation function, differing only by noise realizations. Consequently, all $\mc{S}$ functions except the first vanish.
In this sense, they can be interpreted as response functions that get excited only if the auxiliary field captures genuine clustering information or, equivalently, if it partitions the pair counts in a nontrivial way.
However, having nonzero $\mc{S}_{i>0}$ is only a necessary, but not sufficient, condition for them to encode cosmological information. Ultimately, this depends on how they respond to variations in the cosmological parameters.

\begin{figure}[htbp]
\centering
\includegraphics[width=1\textwidth]{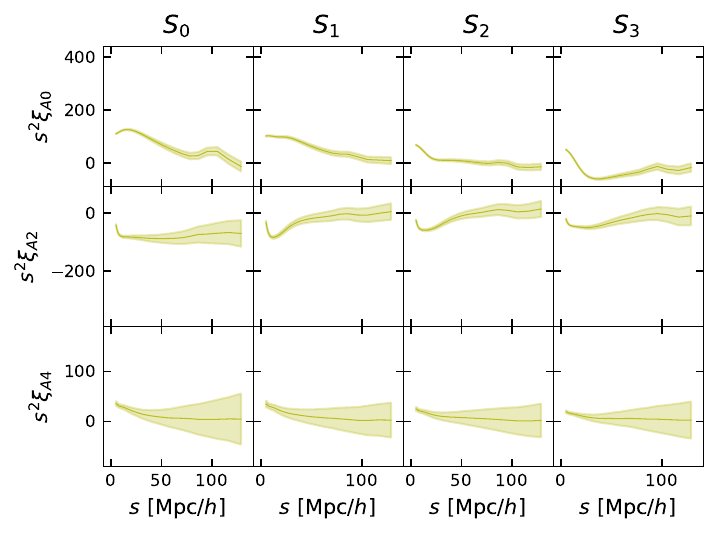}
\caption{Same as Figure~\ref{fig:xi_ebars}, but with the augmented correlation projected onto the Helmert basis.
By construction, the $\mc{S}_0$ column coincides with the standard correlation, corresponding to the right block of Figure~\ref{fig:xi_ebars}.
}
\label{fig:Helmert}
\end{figure}

Figure~\ref{fig:Helmert} illustrates the behaviour of this alternative representation, constructed directly from the $\lambda$-quantiles discussed in Section~\ref{sec:results}.
As expected, $\mathcal{S}_0$ coincides exactly with $\xi$ (compare with the right column of Figure~\ref{fig:xi_ebars}), while the remaining $\mathcal{S}_i$ functions deviate systematically from zero, signalling the presence of additional clustering information.

Since the $\mc{Q}$ and $\mc{S}$ functions are related by a linear transformation, they contain the same information and therefore yield identical Fisher forecasts.
In the context of this work, the $\mc{S}$ representation mainly serves as a complementary visualization and diagnostic tool, providing a convenient way to assess the effectiveness of a given auxiliary field and/or latent variable.
More generally, however, this representation, or a variant of it, may prove advantageous in applications where $\xi$ is modelled separately or where practical computational limitations motivate compressing the quantiles into a reduced set of functions while retaining as much information as possible.

\section{Convergence of Fisher forecasts}\label{app:convergence}

As discussed in Section~\ref{sec:fisher}, Fisher forecasts are inherently idealised, as they rely on simplifying assumptions and only provide a lower bound on the uncertainties expected from real observations.
When these forecasts are derived from numerical simulations, such as the Quijote suite, additional sources of numerical uncertainty must also be considered.
First, the estimation of the covariance matrix is intrinsically noisy because it is based on a finite number of realizations, and its inversion can introduce biases that artificially tighten the inferred parameter constraints.
Similarly, the numerical derivatives entering the Fisher matrix are typically computed from an even smaller number of simulations for each cosmological variation, making them particularly sensitive to sample variance and numerical noise, with the risk of effectively “mistaking noise for signal.”
Finally, although the simulations could in principle be made more representative of realistic galaxy surveys by populating halos with galaxies and including survey geometries and selection effects, they still suffer from intrinsic limitations related to nonlinear gravitational evolution and finite halo resolution, which ultimately constrain their physical realism.
 
For these reasons, the forecasts presented in this work, as well as in similar studies, should be interpreted with caution and viewed primarily as indicative estimates rather than fully realistic predictions.
Nonetheless, the robustness of the results can be partially assessed through a set of standard convergence tests, aimed at quantifying the dependence of the inferred constraints on the number of realizations used in the covariance and derivative estimations, thus addressing the first two sources of uncertainty discussed above.

\begin{figure}[htbp]
\centering
\includegraphics[width=0.39\textwidth]{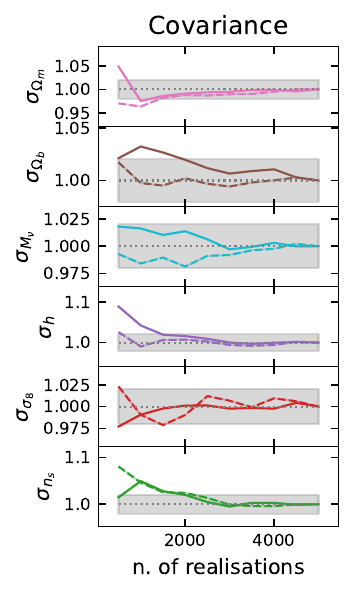}
\includegraphics[width=0.59\textwidth]
{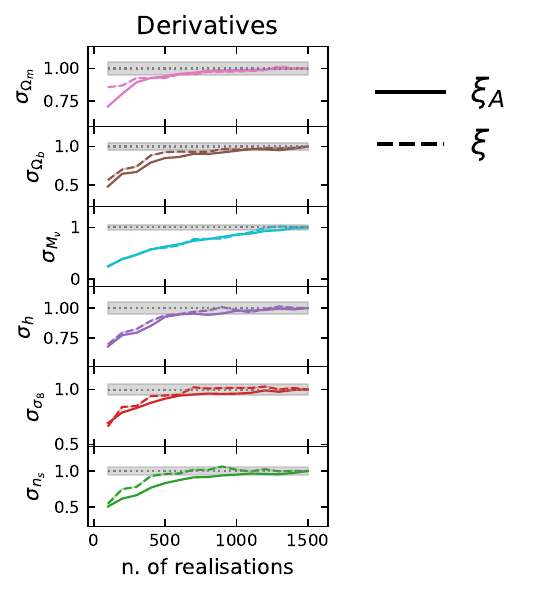}
\caption{Left column: $1\sigma$ Fisher constraints on the cosmological parameters obtained from the augmented (solid) and standard (dashed) correlations as a function of the number of realizations used to estimate their covariance matrices, $\sigma(N_{\mathrm{fid}})$, for the fixed separation range $s \in [10,150],h^{-1}\mathrm{Mpc}$ and $N_q=4$ quantiles.
From top to bottom, the curves correspond to $\Omega_m$ (pink), $\Omega_b$ (brown), $M_{\nu}$ (cyan), $h$ (purple), $\sigma_8$ (red), and $n_s$ (green).
The constraints are normalized to their values at $N_{\mathrm{fid}}=5000$, corresponding to the maximum number of realizations employed, while the gray shaded regions indicate variations at the $2\%$ level.
Central column: same as the left column, but showing the dependence on the number of realizations used to estimate the derivatives, $\sigma(N_{\mathrm{real}})$, with a maximum of $N_{\mathrm{real}}=1500$ realizations.
In this case, the gray shaded regions indicate $5\%$ variations.}
\label{fig:convergence}
\end{figure}

The left panel of Figure~\ref{fig:convergence} illustrates the convergence of the $1\sigma$ parameter constraints with respect to the number of realizations used to estimate the covariance matrix, for both the augmented (solid lines) and standard correlation functions (dashed lines).
The curves are normalized to their values at $N_{\mathrm{fid}} = 5000$, corresponding to the maximum number of realizations employed in this work. 
For $N_{\mathrm{fid}} > 2000$, all parameters remain stable within fluctuations smaller than $2\%$, highlighted by the gray shaded bands, with no evident systematic trend.
This level of stability is consistent with the regular structure of the covariance matrices (Figure~\ref{fig:covmat}), the relatively modest number of bins ($\sim 200$) compared to similar analyses, and the inclusion of the Hartlap correction factor.

The right panel of the figure presents the same convergence test for the numerical derivatives, for which the maximum number of realizations available for each parameter variation is $N_\mathrm{real}=1500$. For $N_\mathrm{real}>1000$, all constraints remain within the $5\%$ variation indicated by the gray shaded band, with the exception of the neutrino mass constraint, which approaches this threshold only at around 1200 realizations.
The figure further indicates that, while the constraints associated with the other cosmological parameters exhibit a clear flattening trend as $N_\mathrm{real}$ increases, the $M_\nu$ curve appears less converged and could benefit from a larger number of realizations.
We also verified that using the alternative finite-difference estimator, $\partial_{M_\nu} \mathbf{d} \simeq \left[-\mathbf{d}(2\Delta M_\nu) +4\,\mathbf{d}(\Delta M_\nu) -3\,\mathbf{d}(M_\nu = 0)
\right] / \left(2\,\Delta M_\nu\right)$, proposed in \cite{villaescusa-navarro2020} produces very similar results.
As an additional consistency check, aimed at ensuring that the signal is not artificially induced by sampling noise, we repeated the analysis after setting to zero all derivatives, for all cosmological parameters, in the separation bins with signal-to-noise ratio smaller than 3.
This procedure did not produce any significant change in the results.

Overall, we conclude that the Fisher forecasts are sufficiently well converged for the purposes of this work, although additional caution is warranted when interpreting the neutrino mass constraints, both in terms of their absolute amplitude and of the improvement factor obtained with the augmented correlation relative to the standard one.

\acknowledgments

We thank I\~nigo S\'{a}ez-Casares for helpful comments on the manuscript.





\bibliographystyle{JHEP}
\bibliography{biblio}





\end{document}